\documentstyle[aps,prl]{revtex}
\input epsf.sty
\def\lsim{\raise1pt\hbox{$<$}\lower3pt\hbox{\llap{$\sim$}}}

\newcommand{\beq}{\begin{equation}}
\newcommand{\eeq}{\end{equation}}
\newcommand{\half}{\mbox{$\textstyle \frac{1}{2}$} }

\newcommand{\ket}[1]{| #1 \rangle}
\newcommand{\bra}[1]{\langle #1 |}
\newcommand{\proj}[1]{\ket{#1}\! \bra{#1}}

\renewcommand{\paragraph}{\section}

\input{psfig}
\begin{document}
\twocolumn[\hsize\textwidth\columnwidth\hsize\csname %
@twocolumnfalse\endcsname

\title
{Capacities of Quantum Erasure Channels}
 
\author{
Charles H. Bennett, David P. DiVincenzo, and
John A. Smolin\\
{\em IBM Research Division, T. J. Watson Research Center,
Yorktown Heights, NY 10598, USA} 
}
 
\date{\today}
\maketitle
 
\begin{abstract}\noindent The quantum analog of the classical erasure
channel provides a simple example of a channel whose asymptotic
capacity for faithful transmission of intact quantum states, with and
without the assistance of a two-way classical side channel, can be
computed exactly. We derive the quantum and classical capacities for
the quantum erasure channel and related channels, and compare them to
the depolarizing channel, for which only upper and
lower bounds on the capacities are known. 
\\Pacs:  03.65.Bz, 42.50.Dv, 89.70.+c
\end{abstract}

] 

Classical information theory, which deals with the optimal use of
classical channels to transmit classical information, has recently
been extended to include the study of quantum channels, and their
optimal use, alone or in conjunction with classical channels, for
communicating not only classical information but also intact quantum
states, and for sharing entanglement between separated observers.  A
classical (discrete, memoryless) channel is generally described by a
set of conditional probabilities $P(j|i)$, the probability of
channel output $j$ given channel input $i$.  A quantum channel may be
described~\cite{Kraus,Schu96} by a trace-preserving, completely
positive linear map (superoperator) ${\cal X}$ from input-state
density matrices to output-state density matrices.  

In classical information theory a channel's capacity is the greatest
asymptotic rate at which classical information can be sent through
the channel
with arbitrarily high reliability. More precisely the capacity (in
bits) of a discrete memoryless channel can be defined as the greatest
number $C$ such that for any rate $R<C$ and any error probability
$\delta>0$, there exist block sizes $m$ and $n$ and an
error-correcting code mapping $m$-bit strings into $n$ forward uses of
the channel with $m/n>R$, such that every $m$-bit string can be
recovered with error probability less than $\delta$ at the receiving
end of the channel.  It is well known that {\em backward\/}
communications, e.g., messages from receiver to sender requesting
retransmission when an error has been detected, do not increase the
forward capacity for classical channels, although they are often used
in practice to reduce latency and complexity of the  decoding
processes.  Another noteworthy feature of classical capacity is that
it is equal to the maximum, over channel input distributions, of the
mutual information between channel input and output for a {\em
single\/} use of the channel.  Thus, the asymptotic capacity for
reliable transmission when the channel is used many times is equal to
the amount of information that can be transmitted unreliably in a
single use of the channel.

\begin{figure}[htbp]
\epsfxsize=8cm
\leavevmode
\epsfbox{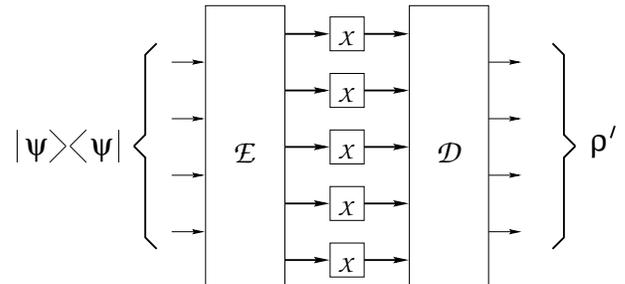}
\caption{A pure input state $\rho=\proj{\psi}$ of $m$
qubits is encoded by a quantum encoder ${\cal E}$ into the
joint state of $n$ intermediate systems, each of which passes
through an independent instance of the quantum channel ${\cal X}$.
The joint state is then decoded by decoder ${\cal D}$ resulting in a
(typically) mixed state $\rho'$ of $m$ qubits, whose fidelity
$F=\bra{\psi}\rho'\ket{\psi}$ relative to the input is evaluated.
This code has a rate $m/n$ of 4/5.}\label{fig1}
\end{figure}

For quantum channels, reliability is measured by {\em fidelity\/} 
\cite{JS,Schu96}, the probability that the channel output would
pass a test for being the same as the input, conducted by someone who
knows what the input was.  When a pure state $\rho=\proj{\psi}$ is sent
into a
quantum channel ${\cal X}$, emerging as an (in general) mixed state
$\rho'={\cal X}(\rho)$,
the fidelity of output relative to the input is 
 \beq F=\bra{\psi}\rho'\ket{\psi}. \eeq \label{fiddef}
Paralleling the definition of capacity for classical channels, the
quantum capacity $Q({\cal X})$ of a quantum channel ${\cal X}$ may be
defined in an asymptotic fashion, as the greatest number $Q$ such that
for any $R<Q$ and any $\delta>0$, there exist block sizes $m$ and $n$
and a quantum error-correcting code mapping states $\ket{\psi}$ of $m$
qubits into $n$ forward uses of the channel with $m/n>R$, such that
any state $\ket{\psi}$ can be recovered with fidelity at least
$1\!-\!\delta$ at the receiving end of the channel.  The encoding and
decoding may be described mathematically as superoperators ${\cal E}$
and ${\cal D}$ on blocks of quantum information carriers, respectively
mapping from $m$ qubits into $n$ intermediate systems (which need not
be qubits), each of which is then sent through an independent instance of 
the  channel, and finally from the $n$ channel
outputs back to $m$ qubits (cf.~Fig.~\ref{fig1}).  Physically a
superoperator corresponds to a unitary interaction of the quantum
system in question with an external system or environment, initially
in a standard pure state.  The superoperator formalism is broad enough
to describe any physically realizable treatment that can be applied to
a quantum system. In particular, mappings between different-sized
Hilbert spaces can be accommodated by adding dummy dimensions to the
smaller space.   This happens explicitly in ${\cal E}$ and ${\cal D}$
and also in channels such as the erasure channels to be described
in this paper.

The above definition of $Q$ is
for a forward quantum channel alone, unassisted by classical
communication.
If we now allow the quantum channel to be assisted by classical
communication, we can define $Q_1$ and $Q_2$ as the asymptotic quantum
capacities of a quantum channel assisted, respectively, by forward and
by two-way classical communication.  We have shown~\cite{BDSW} that
classical forward communication alone does not increase the quantum
capacity of any channel: $Q({\cal X})=Q_1({\cal X})$ for all ${\cal
X}$. Hence $Q$  and $Q_1$ can safely be denoted by a single symbol
$Q$.  By contrast $Q_2$, the quantum capacity assisted by two-way
classical communication, can be greater than $Q$, and is known to be
positive for some channels for which $Q$ is zero. Protocols for
exploiting $Q_2$ typically do not involve a single encoder and
decoder, but rather use multiple adaptive rounds of communication
between the sender and receiver.  The one-way and two-way capacities
$Q$ and $Q_2$ are closely related to the amounts of purified entanglement
distillable, respectively, by one-way and by two-way entanglement
purification protocols from entangled mixed states shared between two
separated observers~\cite{BDSW}.

The three kinds of communication represented by $Q$, $Q_2$, and $C$
differ both fundamentally and practically. The positivity of $Q_2$
determines whether a channel can be used to communicate intact quantum
states and to establish entanglement between separated observers if
reliable storage of quantum information is available.  The positivity
of $Q$ determines whether unreliable quantum storage can be made
reliable, by encoding the data before it is stored and decoding it
after it is retrieved.  The impossibility of sending messages backward
in time precludes two-way protocols in this case. $C$, which we will
now use to denote the
classical capacity of a {\em quantum\/} channel, represents the maximum rate
of classical information transmission allowing arbitrary state
preparations by the sender and arbitrary quantum measurements by the
receiver, including preparations and measurements coherently spanning
multiple information carriers.  

By definition, $Q\leq Q_2$; by using orthogonal quantum states to
transmit classical bits, it follows that $Q\leq C$ for all channels. 
No channels are known for which $Q_2>C$ but we know of no proof that
this is impossible. On the other hand, examples are known (see below) of
channels for which $Q<Q_2$ and for which $Q_2<C$ (cf \cite{BDSW} Sec. VII).
 
The main features of quantum error-correction are illustrated by two
simple channels, operating on a Hilbert space of dimension 2, and
analogous respectively to the classical binary symmetric and binary
erasure channels:
 \begin{itemize}
 \item the {\em depolarizing channel\/} which with probability
$\;\epsilon\;$ replaces the incoming qubit by a qubit in a random state, 
without telling the receiver on which qubits this randomization has been
performed; and
 \item the {\em quantum erasure channel\/} (QEC) \cite{qec}, which with
probability $\;\epsilon\;$ replaces the incoming qubit by an
``erasure state'' $\ket{2}$ orthogonal
to both $\ket{0}$ and $\ket{1}$, thereby both erasing the qubit and
informing the receiver that it has been erased.
 \end{itemize}

Unfortunately, exact expressions are not known for any of the
capacities of the depolarizing channel for general $\;\epsilon,\;$
only upper and lower bounds~\cite{BBPSSW,EM,lafknill,BDSW}. However,
the known bounds are tight enough to show that the depolarizing
channel exhibits the following sequence of thresholds: 

\begin{itemize}
 \item for $\epsilon<0.25408$, all three capacities $Q,Q_2,$ and $C$  
are positive~\cite{BDSW,shorsmolin}.
 \item for $\frac{1}{3}<\epsilon<\frac{2}{3}$, the one-way quantum
capacity $Q$ vanishes but $Q_2$ and $C$ remain
positive~\cite{BBPSSW,EM,lafknill,BDSW}.
 \item for $\frac{2}{3}\leq\epsilon<1$, both quantum capacities vanish
but the classical capacity remains positive~\cite{BDSW}.
 \item at $\epsilon\!=\!1$ (complete depolarization) all capacities vanish.
 \end{itemize}

\begin{figure}[htbp]
\epsfxsize=8.0cm
\leavevmode
\epsfbox{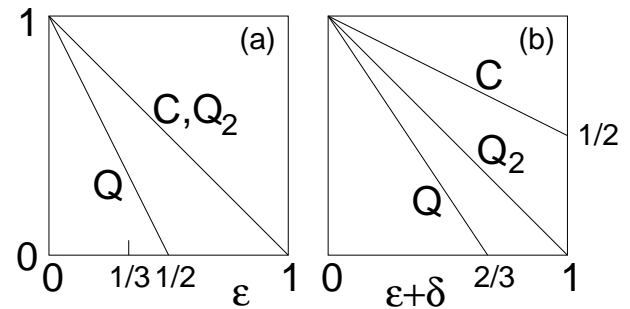}
\caption{(a) Exact classical and quantum capacities for quantum
erasure channel vs.\ erasure probability $\epsilon$. Also
shown is the threshold
$t/n=1/3$ above which quantum codes, in the limit of
large $n$, cannot correct all patterns of $t$ or fewer erasures
in code words of $n$ qubits.
(b) Same capacities for the mixed
erasure/phase-erasure channel with equal probabilities of
erasure ($\epsilon$) and phase erasure ($\delta$) vs.\ total 
erasure probability $\delta+\epsilon$.
}\label{fig2}\end{figure}

The capacities of the QEC, by contrast, can be computed
{\em exactly,\/} (see Fig. \ref{fig2}(a)) and are given by
 \beq\begin{array}{l}
Q= \max\{0,\;1-2\epsilon\} \\
Q_2=C= 1-\epsilon.\\
\end{array}\label{qeccaps}
\eeq

To show that the QEC's one-way capacity $Q$ must vanish
for $\epsilon\geq\half$ suppose the contrary.  The sender (``Alice'')
could then clone quantum information faithfully by dividing it between
two  receivers (e.g.\ ``Bob'' and ``Charlie''), each of whom would think
he was seeing the source through a QEC of
$\epsilon\geq\half$.  In more detail (cf.\ \cite{BDSW}, section IV),
let Alice toss a fair coin for each qubit, and if the result is heads
(resp. tails) send the qubit to Bob (Charlie) through an
$2\epsilon-1$ QEC while sending a pure erasure state to
Charlie (Bob). This implements an $\epsilon\geq\half$ QEC
to each receiver.  Such channels must have zero capacity to prevent cloning.
Linear interpolation between the 
50\% QEC and the noiseless channel\ \cite{cf1}
yields an
upper bound $Q \leq 1-2\epsilon$, which coincides with the lower bound
obtained by using one-way random hash coding
\cite{cf2}.
In such codes, two bits of
redundancy per erased qubit are necessary and sufficient for Bob to
recover the phase and amplitude of all erased qubits with
probability tending to 1 in the limit of large block size.

The QEC's two-way quantum capacity must be at least
$\;1\!-\!\epsilon\;$ by a straightforward construction in which the
sender (``Alice'') uses the QEC, in conjunction with
classical communication, to share $\;1\!-\!\epsilon\;$ good 
Einstein-Podolsky-Rosen (EPR) pairs (such as 
$\frac{1}{\sqrt{2}}\ket{0_A0_B+1_A1_B}$)
with the receiver (``Bob'') per channel use.  These can then be used
to teleport quantum information to Bob at the same rate
$\;1\!-\!\epsilon.\;$ Conversely Alice and Bob could start with an
initial supply of $\;n(1\!-\!\epsilon)\;$ perfectly-entanged EPR
pairs, then use these pairs in conjunction with teleportation to
simulate $n$ instances of a QEC of strength $\epsilon$. If $Q_2$ for
this channel were greater than $\;1\!-\!\epsilon,\;$ Alice and Bob
would have been able to deterministically increase their entanglement
by purely local actions and classical communication, which is
impossible (cf.\ \cite{BDSW}, sect.~II.A).  This establishes that
$Q_2$ is exactly $1-\epsilon$.

Finally, the classical capacity $C$ of the QEC
can be no greater than $\;1\!-\!\epsilon\;$ because of Holevo's upper
bound~\cite{holevo} on the classical capacity of the
$\;1\!-\!\epsilon\;$ non-erased qubits.  Of course
$\;1\!-\!\epsilon\;$ is also the capacity of a {\em classical\/}
erasure channel, which the quantum erasure channel can be made to
simulate by sending in the $\{\ket{0},\ket{1}\}$ basis and receiving
in the $\{\ket{0},\ket{1},\ket{2}\}$ basis.  This establishes that the
classical capacity $C$ of the QEC is exactly $\;1\!-\!\epsilon.$

Another quantum channel for which the capacities can be computed
exactly is the {\em phase erasure channel\/} (PEC), in which, with
probability $\;\epsilon\;$, the phase of the transmitted qubit is
erased without disturbing its amplitude.  This may be described more
formally by a superoperator from one- to two-qubit states, in which
the second output qubit serves as a flag to indicate whether the first
qubit has been subjected to a randomization of its phase.  Thus on an
input $2\times 2$  density matrix $\rho$, the output will be the 
$4\times 4$ density
matrix
 \beq
\rho'=(1\!-\!\epsilon)\;\rho\otimes{\proj{0}}
+\epsilon\;\frac{\rho+\sigma_z\rho\sigma_z^\dagger}{2}\otimes\proj{1}.
 \eeq Here $\sigma_z$ is the diagonal Pauli matrix which introduces a
$\pi$ relative phase between the spin-$z$ eigenstates $\ket{0}$ and
$\ket{1}$ of the first qubit. 
The PEC has unit classical capacity $\;C\!=\!1\;$ for all
$\;\epsilon\;$ because the input states $\ket{0}$ and $\ket{1}$ remain
perfectly distinguishable despite dephasing. The quantum capacities
are $\;Q=Q_2=1-\epsilon\;$ by arguments similar to those given for the
plain erasure channel.  On the one hand, $Q_2$ can be no greater than
$\;1\!-\!\epsilon\;$ because the channel can be simulated by a
noiseless quantum channel of rate $\;1\!-\!\epsilon\;$ supplemented by
classical communication. (Given $n$ qubits, Alice uses the noiseless
channel $\;n(1\!-\!\epsilon)\;$ times to transmit
$\;n(1\!-\!\epsilon)\;$ of the qubits intact, then measures remaining
$n\epsilon$ qubits in the $z$ basis and transmits the results to Bob
classically, allowing him to construct dephased versions of these
qubits.)  On the other hand $\;Q=1\!-\!\epsilon\;$ can be achieved
asymptotically by one-way hash coding~\cite{cf2} because each dephased
qubit contributes one bit of entropy to the syndrome. The no-cloning
argument used to separate $Q$ from $Q_2$ for the QEC does not apply to
the PEC (nor is it needed) because the PEC's preservation of the
amplitude prevents a noiseless quantum channel from being split
into independent PEC's to two or more receivers.

Finally, the QEC and PEC can be generalized to a mixed
erasure/phase-erasure channel that erases qubits with probability
$\;\epsilon\;$ and phase-erases them with probability $\;\delta,\;$
transmitting them undisturbed with probability
$\;1\!-\!\delta\!-\!\epsilon\geq 0$.  By arguments similar to those
already given, the capacities of this channel are (see Fig.~\ref{fig2}(b))
 \beq\begin{array}{lll} 
Q& = & \max\{\;0,\;1-\delta-2\epsilon\;\}. \\
Q_2 &=& 1-\delta-\epsilon \\
C&=& 1-\epsilon
 \end{array}\eeq  The upper bound on $Q$ follows from a slightly more
complex no-cloning argument. Consider a series-parallel combination
which begins with a PEC of strength $\delta$, and is followed by a
parallel combination of a noiseless channel for the phase-erased
qubits and an $\frac{\epsilon}{1-\delta}$-strength QEC for the
non-phase-erased qubits.  When
$\frac{\epsilon}{1-\delta}\geq\frac{1}{2}$, this combination can be
cloned by copying the phase-erased qubits (this introduces no
additional disturbance since dephasing renders quantum data
effectively classical), and splitting the remaining qubits between two
receivers.  Each receiver thus possesses a good copy of all the
dephased qubits and a sufficient number of non-erased, non-dephased
qubits to simulate the erasure part of the channel. For appropriate
values of $\delta$ and $\epsilon$, all three capacities have distinct
nontrivial values in the mixed erasure/phase-erasure channel;
Fig.~\ref{fig2}(b) shows this for the case $\delta=\epsilon$.

It might seem that at least the classical capacity of the depolarizing
channel and other simple channels ought to be known, and indeed that it
should be equal to the maximum classical mutual information that can
be sent through a single use of the channel by optimizing over input
ensembles and output measurements.  In the case of the depolarizing
channel, this one-shot capacity 
 \beq
1\!-\!H_2(\frac{\epsilon}{2})=1+\frac{\epsilon}{2}\log(\frac{\epsilon}{2})+
(1-\frac{\epsilon}{2})\log(1-\frac{\epsilon}{2})
 \eeq\noindent is the capacity of a classical binary symmetric channel
of crossover probability $\epsilon/2$, obtained by using any two
orthogonal states as inputs, and measuring the output in the same
basis. However, we have not been able to rule out the possibility of
achieving a higher capacity by employing input states entangled among
multiple uses of the channel~(cf~\cite{BFS96,holevo96}).  The possibility of
entangled inputs of course does not exist for classical channels, and
their capacity is strictly additive, in the sense that the asymptotic
capacity, as noted previously, is equal to the maximum mutual
information that can be sent through a single use of the channel.

While non-additivity of the classical capacity of quantum channels is
an open question, the quantum capacity $Q$ is definitely known to be
non-additive, in the sense that it sometimes exceeds the maximum {\em
coherent information\/}~\cite{SN96} that can be sent through a single
use of a quantum channel.  Coherent information, which is defined as
the excess of the output state's entropy over the environment's
entropy, is a natural candidate for a measure of distinctively
quantum mutual information because, as Schumacher and Nielsen show~\cite{SN96}, it
cannot be increased by further processing of the channel output,
even with the help of classical communication.
Nonadditivity of quantum capacity is known to occur in particular for
the simple depolarizing channel in the range
$0.25239<\epsilon<0.25408$, where $Q$ is positive but the one-shot coherent
information information is identically zero (by a
25-shot use of the depolarizing channel, \cite{shorsmolin} 
shows the capacity is positive in this range).
The situation is simpler for the QEC, where the maximal
coherent information equals the quantum capacity $Q$ for all $\;\epsilon\;$:
For $\epsilon<\half$, a maximal coherent information equal to $Q$
can be realized by sending a random qubit state into the QEC.  For $\epsilon\geq\half$
it can be realized by sending a fixed qubit, eg $\ket{0},$ into the channel.

A third notion, besides quantum capacity and coherent information,
associated with the ability of channels to transmit intact quantum
states, is the existence of codes able to correct all patterns of $t$
or fewer errors in code words of size $n$.  Rains~\cite{RC} has shown
that, for errors in unknown locations (a situation analogous to the
simple depolarizing channel) such codes cannot exist when $t>(n+1)/6$.
Since a quantum code
can correct $t$ errors at unknown locations iff the same
code can correct $2t$ errors at known locations \cite{qec}, (a situation
analogous to the QEC), there is a range
$1/3\!<\!\epsilon\!<\!1/2$ over which the QEC's capacity $C$
remains positive even though no code can correct all patterns of
$n\epsilon$ erasures in a block of size $n$.  This is possible because
capacity is defined in terms of asymptotically faithful transmission,
which can tolerate some probability of uncorrected errors provided it
tends to zero in the limit of large block size. A similar gap between
perfect and asymptotically faithful error correction occurs for the
QEC's classical capacity $C=1-\epsilon,$ which is strictly
greater the rate of any perfect classical
erasure-correcting code in the limit of large $n$~\cite{Davidsref}.
On the other hand, no gap exists for the QEC's quantum
capacity $Q_2=1-\epsilon$ in the presence of two-way classical
communication.  Here, the teleportation protocol given earlier allows
perfect quantum transmission at a rate $1-t/n$ following any pattern
of $t$ erasures in a block of $n$ qubits.

We thank C. Fuchs and P. W. Shor for helpful discussions at PhysComp
'96 and elsewhere, the Institute for Theoretical Physics for
hospitality (supported by the National Science Foundation under grant
no. PHY94-07194), and the Army Research Office for support.

\begin{thebibliography}{99}
\bibitem{cf1} cf.\ \protect\cite{BDSW}
section V.B where it is shown that capacity cannot be superadditive
for convex combinations of a noisy with a noiseless channel.
\bibitem{cf2}cf.\ \protect\cite{BDSW} section
III.B.3.  These are identical to random linear stabilizer codes;
see \protect\cite{RC} and references therein.
\end{thebibliography}
\end{document}